\documentclass[10pt,conference]{IEEEtran}
\usepackage[utf8]{inputenc}
\usepackage{graphicx}
\usepackage{comment}
\usepackage{booktabs}
\usepackage{float}
\usepackage{adjustbox}
\usepackage[table,xcdraw]{xcolor}
\usepackage{lscape}
 \usepackage{multirow}
 \usepackage[normalem]{ulem}
  \useunder{\uline}{\ul}{}
 \usepackage{lscape}
 \usepackage{longtable}
 \usepackage{array}
 
  \usepackage[normalem]{ulem}
 \useunder{\uline}{\ul}{}
 \usepackage{lscape}
 \usepackage{longtable}

\usepackage{graphicx}
\usepackage{wrapfig}
\usepackage{lscape}
\usepackage{rotating}
\usepackage{epstopdf}

\title{Context-Augmented Software Development Projects: Literature Review and Preliminary Framework}
\author{\IEEEauthorblockN{Glaucia Melo, Paulo Alencar, Don Cowan} 
\IEEEauthorblockA{\textit{David R. Cheriton School of Computer Science} \\
\textit{University of Waterloo}\\
Waterloo, Canada \\
\{gmelo,palencar, dcowan\}@uwaterloo.ca}
}
\begin{document}

\maketitle

\begin{abstract}
Software development is a complex activity which depends on diverse technologies and people's expertise. The approaches to developing software highly depend on these different characteristics, which are the context developers are subject to. This context contains massive knowledge, and not capturing it means knowledge is continuously lost. Although extensively researched, context in software development is still not explicit, nor proposed into a broader view of the context needed by software developers and tools. Therefore, developers' productivity is affected, as the ability to reuse this rich context is hampered. This paper proposes a literature review on context for software development, through nine research questions. The purpose of this study is making the discovered context explicit into an integrated view and proposing a platform to aid software development using context information. We believe supporting contextual knowledge through its representation and mining for recommendation and real-time provision can significantly improve big data software project development. 

\end{abstract}

Keywords: Software engineering, context, adaptive context, software development, literature review.

\section{Introduction}  \label{intro}

Software development in traditional and big data projects is a complex knowledge-intensive \cite{Ciccio2015} effort \cite{MEYER2017}: there is a broad number of different technologies involved in software development. Documentation for the technologies used is everywhere, i.e., Tutorials, Stack Overflow, GitHub, Project Wikis, API Tutorials, and others. Moreover, as a human-centred task \cite{Vasanthapriyan2015}\cite{Ponza2014Prompter}, software development endures diverse practices, based on different software developers' expertise, personal interests, gender \cite{lee2019floss} \cite{imtiaz2019investigating}, how developers manage stress \cite{Sarker2019}, as well as other software development aspects. The approach each software developer take when performing their tasks highly depends on these different characteristics and the context in which developers are. Treating this context as a first-class construct has been pointed out as being essential \cite{Murphy_Beyond2019}, which can lead to transformative changes in how developers can perform their work. 

The information generated or handled by software developers is often massive, and the domains in which software developers work also vary. In addition, software development context is provided in different formats, such as different supporting repositories (e.g., GitHub for code, issue trackers), natural language communication (e.g., code comments, GitHub comments, emails, meeting reports) and many others. Although having recognized its importance demonstrated through the high number of research works in this topic, the substantially variable context in software development is still not explicitly captured, nor proposed as a seed to a unified adaptive framework that can generate recommendations to software developers. In other words, this context is not captured in a more comprehensive model that can be used as a basis to support the development of an adaptive framework to provide in-depth contextual knowledge to software development throughout a project life cycle.

We believe it is essential to understand and capture this context so that adaptive context recommendations about relevant software development information could be provided to developers. The recommendation allows developers to focus on creative tasks other than on how to execute a specific procedural task or wondering what they should do next. Framing this problem as a big data problem brings awareness regarding the possible volume of information that is expected to be dealt with, as well as the velocity of information capture that is necessary. Moreover, the variability of context formats is also an aspect to be considered. Examples of relevant contextual information are the next artifact to be edited or read, an API tutorial, a code snippet, or the knowledge from another developer, of the team or not. To create the grounds for this work, we propose a literature review of context in software development, and through the information collected in this literature review, propose a preliminary context-augmented framework for software development projects.  

To achieve the goal of being able to recommend context in software development projects using an adaptive context framework, we believe we have to pursue the following steps: 
\begin{enumerate}
\item Identify the context: Literature Review (LR) of context in software development;
\item Model the context: propose a model based on the information retrieved from the LR;
\item Propose a preliminary framework that captures software development context and is able to adapt itself to contextual changes.
\end{enumerate}

In this paper, we aim to identify the types of software development context through a Literature Review to lay the foundations for this work and propose a preliminary framework to address the presented problems. The following Research Question is investigated in the literature review: \textit{What types of context have been identified by researchers in software development projects?}. This research question is divided into nine other research questions, detailed in Section \ref{researchmethod}. The information retrieved from the articles is presented in Tables \ref{researchquestions1} and \ref{researchquestions2}.

This paper is structured as follows. Section \ref{intro} presents an introduction to the discussed subject. The Research Method, which grounds the research methodology, presenting a Literature Review, is provided in Section \ref{researchmethod}. The proposed framework is described in Section \ref{adapt}. Conclusions and future work are presented in Section \ref{conclusion}.

\section{Literature Review} \label{researchmethod}

In this section, we present the Literature Review (LR) towards identifying articles that depict contextual information in software development projects. The research method for the literature review is divided into two different steps. First, we performed a search string search, presented in Section \ref{searchstringsearch}. Second, we performed backwards snowballing withing the retrieved papers, presented in Section \ref{snowb}. This section ends with a discussion of the findings in the LR.

\subsection {Search String Search} \label{searchstringsearch}

\subsubsection{Planning Phase} \label{plan}

Literature Reviews are a standard method to obtain evidence on a particular subject, and provides categorized results that have been published in a specific research area \cite{Petersen2015} \cite{Barros2018}. The literature review reported in this paper was conducted to gather the state-of-the-art in the literature regarding current articles that propose research in contextual information for software development, to better understand the use and variability of diverse contexts in software engineering.

\subsubsection{Research Objective and Research Questions} 
The protocol suggested \cite{Petersen2015} uses the GQM approach \cite{VanSolingen2002} to define a goal for a literature review. According to the GQM approach, the goal of this literature review is to: 

\begin{center}
\textbf{analyze} software development 

\textbf{with the purpose of} characterizing 

\textbf{regarding} software developers' context 

\textbf{from the point of view of} researchers 

\textbf{in the context of} software projects
\end{center}

Emerging from the defined objective, the research question this literature review aims to answer is \textbf{\textit{RQ: What types of context have been identified by researchers in software development projects?}}

This question aims to determine the state of practice and lay the foundations for this work. Specifically, we pursue the following characteristics of software development context in the literature:

\textbf{RQ1:} \textit{What are the types or classifications of context?} Is there any type or classification of the context subject of the retrieved article from the literature?

\textbf{RQ2:} \textit{Is there a model specification technique used?} Is there a model specification for the proposed context, for example, an ontology, a model extension or other types of models?

\textbf{RQ3:} \textit{What are the goals or purpose of context?} In this research question, we aim to retrieve the purpose or the goals of the context subject of the retrieved article.

\textbf{RQ4:} \textit{In what step or phase of the software development does the context focus on?} In traditional software development, there are different development phases, such as coding, testing, analysis and others. If a paper identifies the phase where the context can be applied, we want to capture this information and make it explicit.

\textbf{RQ5:} \textit{Is there any evaluation(s) performed?} With this specific research question, we are exploring if any evaluation was performed in the proposed context or the context's purpose.

\textbf{RQ6:} \textit{Are there identified limitations or gaps when using context?} If there are any identified limitations provided within the context proposal or utilization, we also aim to make this information explicit in this literature review.

\textbf{RQ7:} \textit{What are the advantages or disadvantages of this context?} We want to explore the pros and cons of the context retrieved from the literature.

\textbf{RQ8:} \textit{How are the context instances mined} With this research question, we are looking for the uses of the context and if they were mined to retrieve other processed information such, for example, as a recommendation.

\textbf{RQ9:} \textit{Are there any proposed abstractions?} With this research question, we are looking for abstractions of the proposed context within the retrieved article from the literature.

For creating the search string, we have used PICO, proposed by Pai et al. \cite{Pai2004}. The search string is presented in Table \ref{pico}.

\begin{table}[H]
\centering
\caption{Search string creation process with PICO \cite{Pai2004}.}
\begin{tabular}{l}
\hline
\multicolumn{1}{|l|}{(P)opulation: Software developer in software development} \\ \hline
\multicolumn{1}{|l|}{\begin{tabular}[c]{@{}l@{}}Keywords: (Programmer OR (software AND (developer \\ OR tester)) OR ("software development \\ project" OR "software development \\ environment")
\end{tabular}} \\ \hline
 \\ \hline
\multicolumn{1}{|l|}{(I)ntervention Control: Context} \\ \hline
\multicolumn{1}{|l|}{Keywords: (context OR "event based" OR "self adapt")} \\ \hline
 \\ \hline
\multicolumn{1}{|l|}{(C)omparison: None} \\ \hline
 \\ \hline
\multicolumn{1}{|l|}{(O)utcome Measure: Methodology} \\ \hline
\multicolumn{1}{|l|}{\begin{tabular}[c]{@{}l@{}}Keywords: tool* OR system* OR recommend*\end{tabular}} \\ \hline
\end{tabular}
\label{pico}
\end{table}

Having the PICO defined, the search strings for each database is SQ: (Programmer OR (software AND (developer OR tester)) OR ("software development project" OR "software development environment")
AND (Context OR "event based" OR "self adapt" OR skill OR "team size" OR "organizational structure" OR "organizational structure"  OR situational OR "application type" OR "type of application") AND (tool* OR system* OR recommend*).

In terms of article selection, inclusion and exclusion criteria were proposed. These criteria consider articles:

\begin{itemize}
\item Within a Software Engineering scope;
\item About software development;
\item That talk about software development projects;
\item That present studies of context in software development;
\item That are NOT about IoT or hardware.
\end{itemize}

\subsubsection{Execution Phase} \label{executing}

The initial set of articles were retrieved from the ACM Digital Library in August 9th, 2019. The execution phase returned 135 articles. After reading title and abstract, 18 articles were selected for full reading. The complete list of selected articles is presented below. The list shows the year of publication, authors and publication title. 

 \begin{enumerate}

 \item \textbf{1987} - Marzullo, Keith; Wiebe, Douglas - Jasmine: A Software System Modelling Facility;
 \item \textbf{1988} - Alpern, Bowen; Carle, Alan; Rosen, Barry; Sweeney, Peter; Zadeck, Kenneth - Graph Attribution As a Specification Paradigm;
 \item \textbf{1990} - Goldberg, Allen - Reusing Software Developments;
 \item \textbf{1990} - Baker, P. L. - Ada As a Preprocessor Language;
 \item \textbf{2004} - ČubraniĆ, Davor; Murphy, Gail C.; Singer, Janice; Booth, Kellogg S. - Learning from Project History: A Case Study for Software Development;
 \item \textbf{2006} - Mikulovic, Vesna; Heiss, Michael - "How Do I Know What I Have to Do?": The Role of the Inquiry Culture in Requirements Communication for Distributed Software Development Projects;
 \item \textbf{2006} - Rosener, Vincent; Avrilionis, Denis - Elements for the Definition of a Model of Software Engineering;
 \item \textbf{2007} - de Oliveira, Kleber Rocha; de Mesquita Spínola, Mauro - POREI: Patterns-oriented Requirements Elicitation Integrated – Proposal of a Metamodel Patterns-oriented for Integration of the Requirement Elicitation Process;
 \item \textbf{2009} - Ashok, B.; Joy, Joseph; Liang, Hongkang; Rajamani, Sriram K.; Srinivasa, Gopal; Vangala, Vipindeep - DebugAdvisor: A Recommender System for Debugging;
 \item \textbf{2009} - Cataldo, Marcelo; Herbsleb, James D. - End-to-end Features As Meta-entities for Enabling Coordination in Geographically Distributed Software Development;
 \item \textbf{2012} - Tajalli, Hossein; Medvidović, Nenad - A Reference Architecture for Integrated Development and Run-time Environments;
 \item \textbf{2012} - Devos, Nicolas; Ponsard, Christophe; Deprez, Jean-Christophe; Bauvin, Renaud; Moriau, Benedicte; Anckaerts, Guy - Efficient Reuse of Domain-specific Test Knowledge: An Industrial Case in the Smart Card Domain;
 \item \textbf{2013} - Saraiva, Juliana - A Roadmap for Software Maintainability Measurement;
 \item \textbf{2013} - Lin, Jun - Context-aware Task Allocation for Distributed Agile Team;
 \item \textbf{2014} - Wagstrom, Patrick; Datta, Subhajit - Does Latitude Hurt While Longitude Kills? Geographical and Temporal Separation in a Large Scale Software Development Project;
 \item \textbf{2014} - Murphy, Gail C. - Getting to Flow in Software Development;
 \item \textbf{2015} - Lima, Adailton Magalhães; Reis, Rodrigo Quites; Reis, Carla A. Lima - Empirical Evidence of Factors Influencing Project Context in Distributed Software Projects;
 \item \textbf{2019} - Murphy, Gail C. - Beyond Integrated Development Environments: Adding Context to Software Development;
 \end{enumerate}

After fully reading the 18 articles, the articles which had the content according to the defined research questions and inclusion and exclusion criteria are presented in Table \ref{articles_evidence}. The LR is then analyzed in the next subsection. 

\begin{table*}[]
\centering
\caption{Articles retrieved from Search String Search.}
\begin{tabular}{|l|l|l|l|}
\hline
\textbf{ID} & \textbf{Year} & \textbf{Authors} & \textbf{Title} \\ \hline
A1 & 2019 & Murphy,  Gail C. \cite{Murphy_Beyond2019}& Beyond Integrated Development Environments: Adding Context to Software Development \\ \hline
A2 & 2014 & Murphy, Gail C. \cite{Murphy_Getting} & Getting to Flow in Software Development \\ \hline
A3 & 2013 & Lin, Jun \cite{Lin_ctxaware}& Context-aware Task Allocation for Distributed Agile Team \\ \hline
A4 & 2009 & \begin{tabular}[c]{@{}l@{}}Ashok, B.; Joy, Joseph; \\ Liang, Hongkang;  Rajamani, Sriram K.; \\ Srinivasa, Gopal; Vangala, Vipindeep \cite{Ashokdebugadvisor} \end{tabular} & DebugAdvisor: A Recommender System for Debugging \\ \hline
A5 & 2004 & \begin{tabular}[c]{@{}l@{}}ČubraniĆ, Davor; Murphy, Gail C.; \\ Singer, Janice; Booth, Kellogg S. \cite{Cubranic2004Learning}\end{tabular} & Learning from Project History: A Case Study for Software Development \\ \hline
\end{tabular}
\label{articles_evidence}
\end{table*}

\subsubsection{Analysis Phase} \label{analyzing}

In this section, we present the findings of the articles retrieved by the literature review search string method. Each of the nine specific research questions are discussed.

\textbf{RQ1: \textit{What are the types or classifications of the context?}}

Regarding the first research question, the types or classification of context vary. A2, A3 and A5 explicitly mention project task context. These articles define task context as the information around a project task or the relationships in an information space that are relevant to a software developer as they work in a particular task. There is a recent work that has not been indexed yet that also studies project task context \cite{Glaucia2019}. A1 mentions context in a more broad and integrative perspective, by listing the following existing contexts:  Static software structure, Dynamic system execution, Historical artifact changes, Developer activity, and Team and organization activity. The Historical artifact changes proposed by A1 can also relate to the context suggested by A5. A4 proposes context around the error scenario, meaning that the context proposed are natural language text, textual rendering of core dumps or the debugger output of errors that might occur when developing software. 

\textbf{RQ2: \textit{Is there a model specification technique used?}} 

For RQ2, none of the works mention whether there are model specification techniques used for the proposed contexts. 

\textbf{RQ3: \textit{What are the goals or purpose of context?}} 

Regarding RQ3, the objectives of the contexts are clarified. For article A1, each proposed context has a specific goal. The context type is listed below, followed by its purpose. 

\begin{itemize}
\item Static software structure: IDEs provide static source code artifacts as context to tools hosted in the environment.
\item Dynamic system execution: Context in the form of dynamic execution information about a system under development.
\item Historical artifact changes: Tools that access historical information about a system's static artifacts.
\item Developer activity: Context about how humans work to produce the system, and not necessarily what was generated during the system's production. An example is Mylyn's degree of interest. 
\item Team and organization activity: Treating the activities across a value stream as context.
\end{itemize}

A2 states that the use of task context can approximate task context by either capturing developers' interactions or using data from repositories. The context is then used to determine if the information captured or used is relevant to new tasks that will be performed. 

A3 applies task-related information to produce task allocation recommendations. A4 employs error texts to create a query which could be kilobytes of structured and unstructured data containing all contextual data for the issue being debugged. This query allows users to search through all available software repositories (version control, bug database, logs of debugger sessions, etc.). Finally, A5 explains that storing context information (Person, Message, Document, Change Task and File version) can be used to create a project memory from the artifacts and communications created during a software development project's history. Using this context information can facilitate knowledge transfer from experienced to novice developers. 

\textbf{RQ4: \textit{In what step/phase of the software development the context focus?}}

Regarding RQ4, A3, A4 and A5 mention where during the software development, the context should be used. A3 explains its proposed approach should be used during the planning when tasks are being allocated to software developers; A4 states that their proposed context focus on the occurrence of bugs as A5 explains their proposed context can be manipulated during coding or when bugs occur. 

\textbf{RQ5: \textit{Is there any evaluation(s) performed?}} 

Regarding evaluation, subject of RQ5, A1 and A2 do not present any performed evaluation. A3 mentions that a tool was built and evaluated, presenting better results than the tool being used as a comparison. A4 explains the performed evaluations returned useful results (bugs resolutions) for 75\% of the cases tried. Finally, A5 performed a qualitative evaluation regarding the effective use of history information by newcomers within the developed tool that implements the contexts. Results, when tasks are considered complex, are very prominent as \begin{quote}
"the examples of previous changes provided by Hipikat were helpful to newcomers working on the two change tasks. The recommendations were used as pointers to snippets of code that could be reused in the new tasks and as indicators of starting points from which to explore and understand the system. Without
such help, it is hard for a newcomer to a project even to know where to begin." \cite{Cubranic2004Learning}
\end{quote}

\textbf{RQ6: \textit{Are there identified limitations or gaps when using context?}} 

Regarding gaps found, answering RQ6, A1 mention that for the historical artifact changes context type, although many research tools have been  proposed that use historical information, few tools are available to practicing developers. A4 mentions that duplicate bug reports can occur because of code clones, which can hamper evaluation results. None of the other articles mention gaps. 

\textbf{RQ7: \textit{What are the advantages or disadvantages of this context?}} We want to explore the pros and cons of the context retrieved from the literature. 

Regarding RQ7, A1 mention as advantages for the Historical artifact changes the fact that task contexts enable developers to be more productive by making it easy to recall the source code associated with a given task and by allowing other tools, such as content assist, to order information based on work performed as part of the task. As for the Team and organization activity context, the author of the same articles explains that this context enables the correlation of downstream effects with upstream choices and would open new opportunities for feedback to be provided to developers as development
is undertaken. As a disadvantage of this type of context, it is mentioned that this context is still unexplored. A2 mentions as an advantage the fact that a task context can be used to support an interaction style with the increased flow that reduces the information shown to a software developer and enables parts of different information spaces to be related automatically. A3 mentions as an advantage the fact that the tool built to work based on context helps to alleviate a common problem, that is, that tasks were allocated more often to experienced developers, while the less experience ones received a fewer number of tasks to perform. The other works do not mention advantages or disadvantages. 

\textbf{RQ8: \textit{How are the context instances mined?}} 

Regarding RQ8, A3 presents instances as recommendations of who should resolve a task through a tool. A4 suggests as instances the recommendations of information from bug databases considering queries of contextual information about an issue. Finally, A5 describes as instances the recommendation of artifacts that should be edited according to the captured project history. 

\textbf{RQ9: \textit{Are there any abstractions?}} 

For RQ9, no works mention abstractions.

A summary of findings is presented in Tables \ref{researchquestions1} and \ref{researchquestions2}. 

\subsection {Snowballing Search} \label{snowb}

For the area of software engineering context, the term "context" is broad and there are variations in the nomenclature in literature. Therefore, we also implemented snowballing \cite{wohlin2014guidelines}, as an attempt to mitigate this problem with the term context. 

\begin{table*}[]
\centering
\caption{Articles retrieved from Snowballing Search Literature Review.}
\label{snbpapers}
\begin{tabular}{|l|l|l|l|}
\hline
\textbf{ID} & \textbf{Year} & \textbf{Authors}                                                & \textbf{Title}                                                 \\ \hline
SBA1        & 2017          & M. Gasparic, G. C. Murphy, and F. Ricci   \cite{Gasparic_Murphy_Ricci_2017}                      & A context model for IDE-based recommendation systems           \\ \hline
SBA2        & 2018          & N. C. Bradley, T. Fritz, and R. Holmes  \cite{Bradley_Fritz_Holmes_2018}                        & Context-aware Conversational Developer Assistants              \\ \hline
SBA3        & 2003          & D. Čubranić and G. C. Murphy  \cite{Cub_Murphy_2003}                                   & Hipikat: Recommending Pertinent Software Development Artifacts \\ \hline
SBA4        & 2014          & L. Ponzanelli, G. Bavota, M. D. Penta, R. Oliveto, and M. Lanza \cite{Ponzanelli_Bavota_Penta_Oliveto_Lanza_2014} & Prompter: A Self-Confident Recommender System                  \\ \hline
SBA5        & 2007          & F. W. Warr and M. P. Robillard \cite{Warr_Robillard_2007}                                 & Suade: Topology-Based Searches for Software Investigation      \\ \hline
SBA6        & 2005          & R. Holmes and G. C. Murphy \cite{Holmes_Murphy_2005}                                     & Using Structural Context to Recommend Source Code Examples     \\ \hline
SBA7        & 2006          & M. Kersten and G. C. Murphy  \cite{Kersten_Murphy_2006}                                   & Using Task Context to Improve Programmer Productivity          \\ \hline
\end{tabular}
\end{table*}

The papers retrieved from the literature review were used as seed for the snowballing literature study.  From the seed papers, we have performed a backward snowballing search step \cite{jalali2012systematic}, i.e., we have looked at all their references, going backward in the citation graph. We stopped the process within the first set of collected papers. After we selected the papers, we have also collected information regarding the research questions from these papers. 

For the backward snowballing (reference search), 87 papers were extracted from the references from the five seed papers. After deleting the duplicates, 81 papers were left. The title and abstract of each paper was read, looking for papers according to the objectives set in Section \ref{plan}. After title and abstract exclusion, 14 papers were selected for full reading. Duplicate publications about the same solution in different proceeding or transactions were excluded. After fully reading the articles, 7 papers were selected for analysis, according to the same inclusion and exclusion criteria used during the Search String Search step. These papers are presented in Table \ref{snbpapers}.

These papers were fully read and relevant information about each of them was included in  Table \ref{researchquestions1} and Table \ref{researchquestions2}. These articles are identified by the ID "SBAx" in the table. 

\subsection{Further Discussion} 

We have noticed that the first round of articles had articles from 1977 until 2019. However, the selected articles ranged from 2003 to 2019. The majority of the articles use context as the information around project tasks and the information about the tasks. One article also considers the information from errors raised during software development. All the instances of the contexts mentioned were aiming at recommending information to software developers. The advantages and disadvantages were very specific to each proposal. We could also perceive that the contexts of the articles analyzed are within different steps of the software development cycle, although using the same information (project task context), from which it can be infered that a platform can be built for more than one step of the software development exploring multiple purposes of the same context information. 

\section{Adaptive Context-Augmented Framework for Software Development Projects} \label{adapt}

\subsection{The Big Picture}
 
Due to the variability of context observed in the literature review (environment, people, domain), we propose a framework to capture the software development context, monitor the possible variabilities and recommend to developers specific knowledge and potential next steps.

Context can be defined as something that is part of an environment and can be sensed. A more specific definition applied to software engineering, proposed by Murphy \cite{Murphy_Beyond2019} is that it "is the information about the system under development and the environment and process in which the system is being developed". A system that can respond to these possible mutable scenarios (e.g., domain, process, technologies involved, people) beats methods that are not prepared for these contextual changes \cite{nascimento2018iot}. 

\begin{figure}[h]
 \centering
 \includegraphics[width=0.5\textwidth]{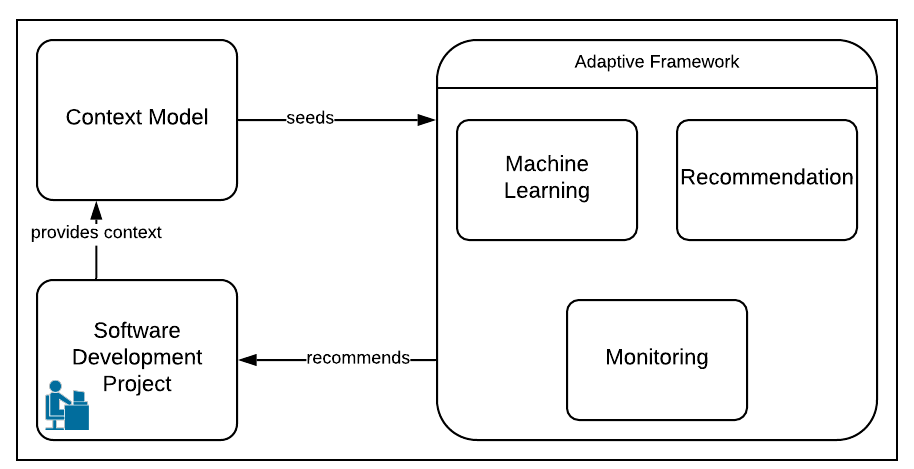}
 \caption{Proposed high-level adaptive context-augmented framework for software development projects.}
 \label{fig:adaptiveframw}
\end{figure}

As we aim to develop an adaptive context framework, we propose a framework based on the observed context variability. Therefore, the following modules are proposed: (i) the software development project where the context model will be applied to; (ii) a baseline of a reconfigurable context model; and (iii) an engine that adapts machine learning models to the context model and provides recommendations to software developers. A high-level framework representation is shown in Figure \ref{fig:adaptiveframw}.

An example of a possible context model, according to the information retrieved in the literature review is presented next.

\subsection{Context Model Example}

\begin{figure}[h]
 \centering
 \includegraphics[width=0.5\textwidth]{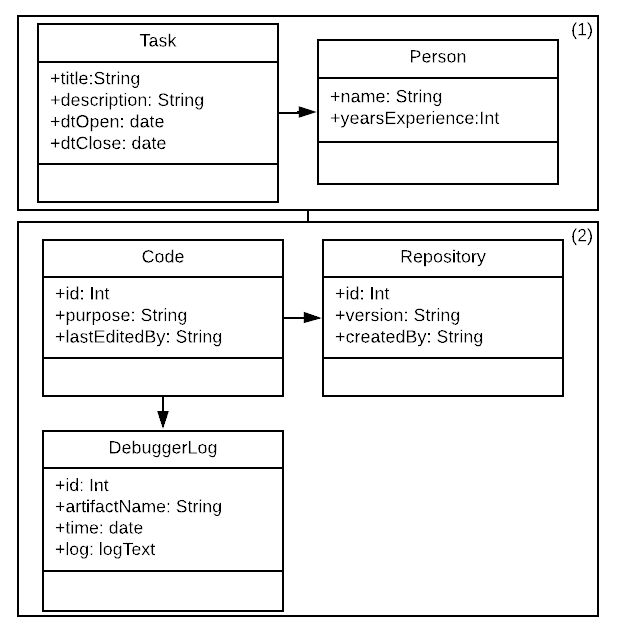}
 \caption{Context Model.}
 \label{fig:contextmodel}
\end{figure}

This Context Model proposed as part of the Adaptive Context-Augmented Framework in Figure \ref{fig:adaptiveframw} is based on the context types (RQ1) identified in the Literature Review (Section \ref{researchmethod}). We do not claim this model is final; we use the model as a basis to understand some of the possible contextual variabilities and to guide us to produce practical examples. The Context Model is presented in Figure \ref{fig:contextmodel}.

An illustrative example on how this model can be used and integrated with the proposed framework is presented next.  

\subsection{Illustrative Example}

Gabi is a software developer that has been working with Java programming for nine years. She has been recently working in project X, a new project of the company. When there is a new project, and Gabi needs to create an MVP to show her clients, she deploys the software locally, using a container tool such as Docker and manually uploading the project to a web server. She also reboots the server manually after each deployment, so changes are effective. This way is faster, and she does not have to configure a job or a server to generate a deployment automatically, which would take her long and the cause the clients to wait much longer for the MVP. 

In project Y, a mature and huge project in the company, when a version of the software has to go to production, all Gabi does is to commit the code from the local to the shared code repository. Then, the scheduled automated job in a Jenkins server will take care of the other steps, which are checking out the code, building the project, uploading the package on the server and rebooting the server.

\begin{figure}[H]
 \centering
 \includegraphics[width=0.5\textwidth]{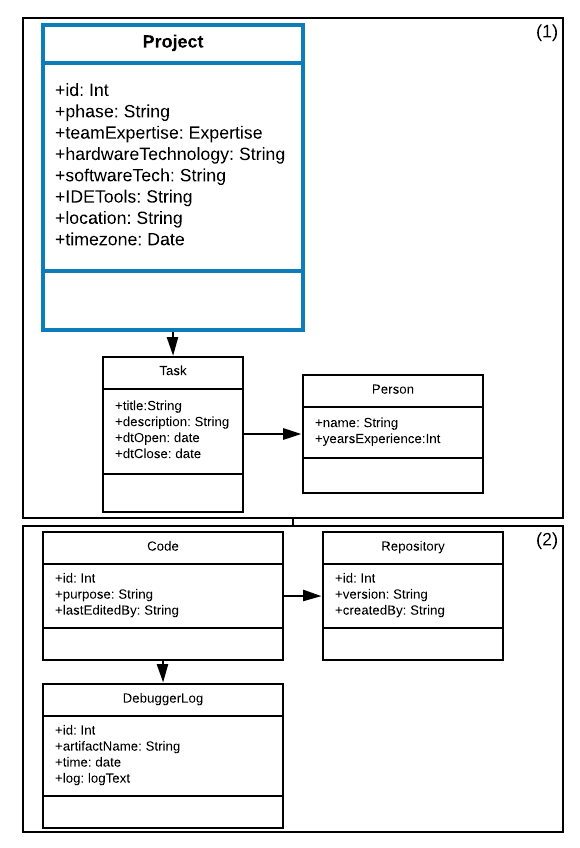}
 \caption{Extended Context Model.}
 \label{fig:contextmodelnew}
\end{figure}

In theory, the steps are the same, but because the projects are different, Gabi's work is different, which means that in the second case, the context model should be expecting Project information or project phase information (MVP / in Production), which defines how the deployment will be done. If Gabi, who has been working in project Y for years, forgets she needs to deploy Project X manually, this can be a problem. In this case, the context model should be adapted to recognize and store the project identification context (including the phase/maturity of the project), and should be as shown in Figure \ref{fig:contextmodelnew}. This figure also shows several other contextual Project attributes, including team expertise, hardware and software technologies, IDE tools, as well as location and timezone, that can also influence Gabi's work.  

\section{Conclusions} \label{conclusion}

The purpose of the current study was to determine how software development context is captured in the literature and propose a preliminary context model that can serve as a foundation to support the identified context. These are the first steps to a solution that explicitly considers software development context to provide in-depth contextual knowledge to software projects throughout a project life cycle. Prior studies have noted the importance of the presence or absence of context information, and how it can influence recommendations during software development \cite{Glaucia2019}. The proposed framework should then be adaptive to the contextual factors that are available and provide recommendations accordingly. 

Although context for software development is a subject explored through research, no work recognizes context broadly, unitedly or as a big data problem. With the results of this work, we hope to provide a basis for future reference and research. Future work involves gathering datasets to add more context information (if new context should be added) and further investigating the challenges that can arise from this proposal.

\bibliographystyle{IEEEtran} 
\bibliography{main}

\begin{thebibliography}{10}
\providecommand{\url}[1]{#1}
\csname url@samestyle\endcsname
\providecommand{\newblock}{\relax}
\providecommand{\bibinfo}[2]{#2}
\providecommand{\BIBentrySTDinterwordspacing}{\spaceskip=0pt\relax}
\providecommand{\BIBentryALTinterwordstretchfactor}{4}
\providecommand{\BIBentryALTinterwordspacing}{\spaceskip=\fontdimen2\font plus
\BIBentryALTinterwordstretchfactor\fontdimen3\font minus
  \fontdimen4\font\relax}
\providecommand{\BIBforeignlanguage}[2]{{%
\expandafter\ifx\csname l@#1\endcsname\relax
\typeout{** WARNING: IEEEtran.bst: No hyphenation pattern has been}%
\typeout{** loaded for the language `#1'. Using the pattern for}%
\typeout{** the default language instead.}%
\else
\language=\csname l@#1\endcsname
\fi
#2}}
\providecommand{\BIBdecl}{\relax}
\BIBdecl

\bibitem{Ciccio2015}
C.~Di~Ciccio, A.~Marrella, and A.~Russo,
  ``\BIBforeignlanguage{English}{Knowledge-intensive processes:
  Characteristics, requirements and analysis of contemporary approaches},''
  \emph{\BIBforeignlanguage{English}{Journal on Data Semantics}}, vol.~4,
  no.~1, pp. 29--57, Mar 2015.

\bibitem{MEYER2017}
\BIBentryALTinterwordspacing
A.~N. Meyer, L.~E. Barton, G.~C. Murphy, T.~Zimmermann, and T.~Fritz,
  ``\BIBforeignlanguage{English}{The work life of developers: Activities,
  switches and perceived productivity},''
  \emph{\BIBforeignlanguage{English}{IEEE Transactions on Software
  Engineering}}, vol.~43, no.~12, pp. 1178--1193, 2017. [Online]. Available:
  \url{http://ieeexplore.ieee.org/document/7829407}
\BIBentrySTDinterwordspacing

\bibitem{Vasanthapriyan2015}
S.~Vasanthapriyan, J.~Tian, and J.~Xiang, ``A survey on knowledge management in
  software engineering,'' in \emph{Software Quality, Reliability and
  Security-Companion (QRS-C), 2015 IEEE International Conference on}.\hskip 1em
  plus 0.5em minus 0.4em\relax IEEE, 2015, pp. 237--244.

\bibitem{Ponza2014Prompter}
\BIBentryALTinterwordspacing
L.~Ponzanelli, G.~Bavota, M.~Di~Penta, R.~Oliveto, and M.~Lanza,
  ``\BIBforeignlanguage{English}{Mining stackoverflow to turn the ide into a
  self-confident programming prompter}.''\hskip 1em plus 0.5em minus
  0.4em\relax ACM, May 31, 2014, pp. 102--111. [Online]. Available:
  \url{http://dl.acm.org/citation.cfm?id=2597077}
\BIBentrySTDinterwordspacing

\bibitem{lee2019floss}
A.~Lee and J.~C. Carver, ``Floss participants' perceptions about gender and
  inclusiveness: a survey,'' in \emph{Proceedings of the 41st International
  Conference on Software Engineering}.\hskip 1em plus 0.5em minus 0.4em\relax
  IEEE Press, 2019, pp. 677--687.

\bibitem{imtiaz2019investigating}
N.~Imtiaz, J.~Middleton, J.~Chakraborty, N.~Robson, G.~Bai, and E.~Murphy-Hill,
  ``Investigating the effects of gender bias on github,'' in \emph{Proceedings
  of the 41st International Conference on Software Engineering}.\hskip 1em plus
  0.5em minus 0.4em\relax IEEE Press, 2019, pp. 700--711.

\bibitem{Sarker2019}
F.~Sarker, B.~Vasilescu, K.~Blincoe, and V.~Filkov, ``Socio-technical work-rate
  increase associates with changes in work patterns in online projects,'' in
  \emph{Proceedings of the 41st International Conference on Software
  Engineering}.\hskip 1em plus 0.5em minus 0.4em\relax IEEE Press, 2019, pp.
  936--947.

\bibitem{Murphy_Beyond2019}
G.~Murphy, ``Beyond integrated development environments: adding context to
  software development,'' in \emph{Proceedings of the 41st International
  Conference on Software Engineering}.\hskip 1em plus 0.5em minus 0.4em\relax
  IEEE Press, 2019, pp. 73--76.

\bibitem{Petersen2015}
K.~Petersen, S.~Vakkalanka, and L.~Kuzniarz, ``Guidelines for conducting
  systematic mapping studies in software engineering: An update,''
  \emph{Information and Software Technology}, vol.~64, pp. 1--18, 2015.

\bibitem{Barros2018}
\BIBentryALTinterwordspacing
J.~L. Barros-Justo, F.~B.~V. Benitti, and A.~L. Cravero-Leal, ``Software
  patterns and requirements engineering activities in real-world settings: A
  systematic mapping study,'' pp. 23--42, 2018, iD: 271914. [Online].
  Available:
  \url{http://www-sciencedirect-com.ez29.capes.proxy.ufrj.br/science/article/pii/S0920548917303173}
\BIBentrySTDinterwordspacing

\bibitem{VanSolingen2002}
R.~Van~Solingen, V.~Basili, G.~Caldiera, and H.~D. Rombach, ``Goal question
  metric (gqm) approach,'' \emph{Encyclopedia of software engineering}, 2002.

\bibitem{Pai2004}
M.~Pai, M.~McCulloch, J.~D. Gorman, N.~Pai, W.~Enanoria, G.~Kennedy,
  P.~Tharyan, and J.~J. Colford, ``Systematic reviews and meta-analyses: an
  illustrated, step-by-step guide.'' \emph{The National medical journal of
  India}, vol.~17, no.~2, pp. 86--95, 2004, pmid:15141602.

\bibitem{Murphy_Getting}
\BIBentryALTinterwordspacing
G.~C. Murphy, ``Getting to flow in software development,'' in \emph{Proceedings
  of the 2014 ACM International Symposium on New Ideas, New Paradigms, and
  Reflections on Programming \& Software}, ser. Onward! 2014.\hskip 1em plus
  0.5em minus 0.4em\relax New York, NY, USA: ACM, 2014, pp. 269--281. [Online].
  Available: \url{http://doi.acm.org/10.1145/2661136.2661158}
\BIBentrySTDinterwordspacing

\bibitem{Lin_ctxaware}
\BIBentryALTinterwordspacing
J.~Lin, ``Context-aware task allocation for distributed agile team,'' in
  \emph{Proceedings of the 28th IEEE/ACM International Conference on Automated
  Software Engineering}, ser. ASE'13.\hskip 1em plus 0.5em minus 0.4em\relax
  Piscataway, NJ, USA: IEEE Press, 2013, pp. 758--761. [Online]. Available:
  \url{https://doi.org/10.1109/ASE.2013.6693151}
\BIBentrySTDinterwordspacing

\bibitem{Ashokdebugadvisor}
\BIBentryALTinterwordspacing
B.~Ashok, J.~Joy, H.~Liang, S.~K. Rajamani, G.~Srinivasa, and V.~Vangala,
  ``Debugadvisor: A recommender system for debugging,'' in \emph{Proceedings of
  the 7th Joint Meeting of the European Software Engineering Conference and the
  ACM SIGSOFT Symposium on The Foundations of Software Engineering}, ser.
  ESEC/FSE '09.\hskip 1em plus 0.5em minus 0.4em\relax New York, NY, USA: ACM,
  2009, pp. 373--382. [Online]. Available:
  \url{http://doi.acm.org/10.1145/1595696.1595766}
\BIBentrySTDinterwordspacing

\bibitem{Cubranic2004Learning}
\BIBentryALTinterwordspacing
D.~\v{C}ubrani\'{C}, G.~C. Murphy, J.~Singer, and K.~S. Booth, ``Learning from
  project history: A case study for software development,'' in
  \emph{Proceedings of the 2004 ACM Conference on Computer Supported
  Cooperative Work}, ser. CSCW '04.\hskip 1em plus 0.5em minus 0.4em\relax New
  York, NY, USA: ACM, 2004, pp. 82--91. [Online]. Available:
  \url{http://doi.acm.org/10.1145/1031607.1031622}
\BIBentrySTDinterwordspacing

\bibitem{Glaucia2019}
G.~Melo, T.~Oliveira, P.~Alencar, and D.~Cowan,
  ``\BIBforeignlanguage{English}{Retrieving curated stack overflow posts from
  project task similarities},'' in
  \emph{\BIBforeignlanguage{English}{International Conference on Software
  Engineering Knowledge Engineering}}, 2019, pp. 415--418.

\bibitem{wohlin2014guidelines}
C.~Wohlin, ``Guidelines for snowballing in systematic literature studies and a
  replication in software engineering,'' in \emph{Proceedings of the 18th
  international conference on evaluation and assessment in software
  engineering}.\hskip 1em plus 0.5em minus 0.4em\relax Citeseer, 2014, p.~38.

\bibitem{Gasparic_Murphy_Ricci_2017}
M.~Gasparic, G.~C. Murphy, and F.~Ricci, ``A context model for ide-based
  recommendation systems,'' \emph{Journal of Systems and Software}, vol. 128,
  p. 200–219, Jun 2017.

\bibitem{Bradley_Fritz_Holmes_2018}
\BIBentryALTinterwordspacing
N.~C. Bradley, T.~Fritz, and R.~Holmes, ``Context-aware conversational
  developer assistants,'' in \emph{Proceedings of the 40th International
  Conference on Software Engineering}, ser. ICSE ’18.\hskip 1em plus 0.5em
  minus 0.4em\relax ACM, 2018, p. 993–1003, event-place: Gothenburg, Sweden.
  [Online]. Available: \url{http://doi.acm.org/10.1145/3180155.3180238}
\BIBentrySTDinterwordspacing

\bibitem{Cub_Murphy_2003}
\BIBentryALTinterwordspacing
D.~Čubranić and G.~C. Murphy, ``Hipikat: Recommending pertinent software
  development artifacts,'' in \emph{Proceedings of the 25th International
  Conference on Software Engineering}, ser. ICSE ’03.\hskip 1em plus 0.5em
  minus 0.4em\relax IEEE Computer Society, 2003, p. 408–418, event-place:
  Portland, Oregon. [Online]. Available:
  \url{http://dl.acm.org/citation.cfm?id=776816.776866}
\BIBentrySTDinterwordspacing

\bibitem{Ponzanelli_Bavota_Penta_Oliveto_Lanza_2014}
L.~Ponzanelli, G.~Bavota, M.~D. Penta, R.~Oliveto, and M.~Lanza, ``Prompter: A
  self-confident recommender system,'' in \emph{2014 IEEE International
  Conference on Software Maintenance and Evolution}, Sep 2014, p. 577–580.

\bibitem{Warr_Robillard_2007}
\BIBentryALTinterwordspacing
F.~W. Warr and M.~P. Robillard, ``Suade: Topology-based searches for software
  investigation,'' in \emph{Proceedings of the 29th International Conference on
  Software Engineering}, ser. ICSE ’07.\hskip 1em plus 0.5em minus
  0.4em\relax IEEE Computer Society, 2007, p. 780–783. [Online]. Available:
  \url{https://doi.org/10.1109/ICSE.2007.80}
\BIBentrySTDinterwordspacing

\bibitem{Holmes_Murphy_2005}
\BIBentryALTinterwordspacing
R.~Holmes and G.~C. Murphy, ``Using structural context to recommend source code
  examples,'' in \emph{Proceedings of the 27th International Conference on
  Software Engineering}, ser. ICSE ’05.\hskip 1em plus 0.5em minus
  0.4em\relax ACM, 2005, p. 117–125, event-place: St. Louis, MO, USA.
  [Online]. Available: \url{http://doi.acm.org/10.1145/1062455.1062491}
\BIBentrySTDinterwordspacing

\bibitem{Kersten_Murphy_2006}
\BIBentryALTinterwordspacing
M.~Kersten and G.~C. Murphy, ``Using task context to improve programmer
  productivity,'' in \emph{Proceedings of the 14th ACM SIGSOFT International
  Symposium on Foundations of Software Engineering}, ser. SIGSOFT
  ’06/FSE-14.\hskip 1em plus 0.5em minus 0.4em\relax ACM, 2006, p. 1–11.
  [Online]. Available: \url{http://doi.acm.org/10.1145/1181775.1181777}
\BIBentrySTDinterwordspacing

\bibitem{jalali2012systematic}
S.~Jalali and C.~Wohlin, ``Systematic literature studies: database searches vs.
  backward snowballing,'' in \emph{Proceedings of the 2012 ACM-IEEE
  international symposium on empirical software engineering and
  measurement}.\hskip 1em plus 0.5em minus 0.4em\relax IEEE, 2012, pp. 29--38.

\bibitem{nascimento2018iot}
N.~Nascimento, P.~Alencar, C.~Lucena, and D.~Cowan, ``An iot analytics embodied
  agent model based on context-aware machine learning,'' in \emph{2018 IEEE
  International Conference on Big Data (Big Data)}.\hskip 1em plus 0.5em minus
  0.4em\relax IEEE, 2018, pp. 5170--5175.

\end{thebibliography}

\begin{landscape}
\begin{table}[]
\centering
\caption{RQs Table Summary - RQ1 to RQ4}
\label{researchquestions1}
\begin{tabular}{|l|l|l|l|l|}
\hline
\textbf{ID} & \textbf{RQ1} & \textbf{RQ2} & \textbf{RQ3} & \textbf{RQ4} \\ \hline
\textbf{A1} & Static software structure &  & \begin{tabular}[c]{@{}l@{}}IDEs provide static source code artifacts \\ as context to tools hosted in the environment\end{tabular} &  \\ \hline
\textbf{} & Dynamic system execution &  & \begin{tabular}[c]{@{}l@{}}Context in the form of dynamic execution \\ information about a system under development.\end{tabular} &  \\ \hline
\textbf{} & Historical artifact changes &  & \begin{tabular}[c]{@{}l@{}}Tools that access historical informatio\\ nabout a system’s static artifacts.\end{tabular} &  \\ \hline
\textbf{} & Developer activity &  & \begin{tabular}[c]{@{}l@{}}Context about how humans work to produce the system, \\ and not necessarily what was generated during the \\ system's production. Ex.: Mylyn's degree of interest.\end{tabular} &  \\ \hline
\textbf{} & Team and organization activity &  & Treating the activities across a value streamas context. &  \\ \hline
\textbf{A2} & \begin{tabular}[c]{@{}l@{}}Task Context: pieces and relationships\\  in an information space (e.g., artifacts) \\ that are relevant to a software developer\\  as they work on a particular task\end{tabular} &  & \begin{tabular}[c]{@{}l@{}}Approximate task context by either capturing developers' \\ interactions or using data from repositories to determine \\ if the information captured or used is relevant to new \\ tasks that will be performed.\end{tabular} &  \\ \hline
\textbf{A3} & \begin{tabular}[c]{@{}l@{}}Approach uses current task related \\ information generated during sprint\\  assessment phase (or in sprint planning\\   meeting), task completion information  \\ generated during previous sprint review \\ phase/meeting, and characteristics of team\\  members.\end{tabular} &  & Produce task allocations recommendations. & Planning (task allocation) \\ \hline
\textbf{A4} & \begin{tabular}[c]{@{}l@{}}Natural language text, textual rendering \\ of core dumps, debugger output etc.\end{tabular} &  & \begin{tabular}[c]{@{}l@{}}Have fat query, a query which could be kilobytes of \\ structured and unstructured data containing all contextual\\  information for the issue being debugged. Allows users to \\ search through all available software repositories (version\\  control, bug database, logs of debugger sessions, etc).\end{tabular} & When bug occurs \\ \hline
\textbf{A5} & \begin{tabular}[c]{@{}l@{}}Person, Message, Document, Change \\ Task and File version.\end{tabular} &  & \begin{tabular}[c]{@{}l@{}}Create a project memory from the artifacts and communications \\ created during a software development project’s history \\ to facilitate knowledge transfer from experienced \\ developers to novice.\end{tabular} & Coding or when bug occurs. \\ \hline
\textbf{SBA1} & \begin{tabular}[c]{@{}l@{}}13 IDE contexts. The authors list each and \\ characterize into the following categories: \\ who, what, when and where.\end{tabular} & \begin{tabular}[c]{@{}l@{}}Yes. Contextual \\ factors of interactions \\ of developers with IDEs\end{tabular} & \begin{tabular}[c]{@{}l@{}}Support development in an IDE and support context-aware\\  RSSE systems development.\end{tabular} & \begin{tabular}[c]{@{}l@{}}Mostly coding and testing, \\ which are done in an IDE such as Eclipse.\end{tabular} \\ \hline
\textbf{SBA2} & \begin{tabular}[c]{@{}l@{}}12 elements are modeled. They are \\ related to the current file, local\\  repository, remote \\ repository and other \\ services such as test and assign\\  a reviewer to the code.\end{tabular} & \begin{tabular}[c]{@{}l@{}}Yes. The contexts are \\ modeled so the \\ conversational assistant \\ can use the model to \\ retrieve the responses.\end{tabular} & \begin{tabular}[c]{@{}l@{}}Reducing low-level commands that developers need to perform,\\  freeing them to focus on their high-level tasks through voice \\ commands.\end{tabular} & Coding and assigning test. \\ \hline
\textbf{SBA3} & \begin{tabular}[c]{@{}l@{}}Already mapped during first step of the LR \\ in paper Learning from project history.\end{tabular} &  &  &  \\ \hline
\textbf{SBA4} & Code &  & Retrieve Stack Overflow Posts according to code typed in IDE & Coding and assigning test. \\ \hline
\textbf{SBA5} & \begin{tabular}[c]{@{}l@{}}Methods or fields during programming that \\ users specify as relevant (often result of a \\ text search)\end{tabular} &  & \begin{tabular}[c]{@{}l@{}}Help developers quickly find relevant elements and understand \\ their relationships with the other elements  that implement the \\ feature of interest\end{tabular} & Coding \\ \hline
\textbf{SBA6} & Code, Methods and Field Declaration &  & \begin{tabular}[c]{@{}l@{}}Support developers to locate relevant code to the \\ current code being written\end{tabular} & Coding \\ \hline
\textbf{SBA7} & Iteraction events history of a task & Task context model proposed. & \begin{tabular}[c]{@{}l@{}}The proposed model reduces information overload \\ and focuses on a programmer’s work by filtering and \\ ranking the information presented \\ by the development environment\end{tabular} & Coding \\ \hline
\end{tabular}
\end{table}
\end{landscape}

\begin{landscape}
\begin{table}[]
\centering
\caption{RQs Table Summary - RQ5 to RQ9}
\label{researchquestions2}
\begin{tabular}{|l|l|l|l|l|l|}
\hline
\textbf{ID} & \textbf{RQ5} & \textbf{RQ6} & \textbf{RQ7} & \textbf{RQ8} & \textbf{RQ9} \\ \hline
\textbf{A1} &  &  &  &  &  \\ \hline
\textbf{} &  &  &  &  &  \\ \hline
\textbf{} &  & \begin{tabular}[c]{@{}l@{}}Although many research\\ tools have been  \\ proposed that use \\ historical information, \\ few tools are available \\ to practicing developers.\end{tabular} & \begin{tabular}[c]{@{}l@{}}Adv: Task contexts enable developers to be \\ more productive by making it easy to recall \\ the source code associated with a given task \\ and by enabling other tools, such as content \\ assist, to order  information based on work \\ performed as part of the task.\end{tabular} &  &  \\ \hline
\textbf{} &  &  &  &  &  \\ \hline
\textbf{} &  &  & \begin{tabular}[c]{@{}l@{}}Adv: Enables correlation of downstream effects \\ with upstream choices and would open new \\ opportunities for feedback to be provided to \\ developers. Disad.: Still unexplored.\end{tabular} &  &  \\ \hline
\textbf{A2} &  &  & \begin{tabular}[c]{@{}l@{}}Adv: Increased flow reduces information \\ shown to developers and enables different \\ information to be related automatically.\end{tabular} &  &  \\ \hline
\textbf{A3} & Tool presented better results than baseline. &  & \begin{tabular}[c]{@{}l@{}}The tool also balances a common problem \\ of the increased number of tasks allocated  to \\ experienced users, while others were idle.\end{tabular} & \begin{tabular}[c]{@{}l@{}}Recommendation of who should \\ resolve a task through a tool.\end{tabular} &  \\ \hline
\textbf{A4} & \begin{tabular}[c]{@{}l@{}}One of the performed evaluations returned \\ useful results (bugs resolutions) for 75\% of \\ the cases tried.\end{tabular} & \begin{tabular}[c]{@{}l@{}}Duplicate bug reports \\ because of code clones, \\ which can hamper evaluation \\ results.\end{tabular} &  & \begin{tabular}[c]{@{}l@{}}Recommendations of information \\ from bug databases considering \\ query of contextual information \\ of issue.\end{tabular} &  \\ \hline
\textbf{A5} & \begin{tabular}[c]{@{}l@{}}Qualitative evaluation regarding effective use \\ of history information by newcomers within \\ developed tool that implements the contexts.\end{tabular} &  &  & \begin{tabular}[c]{@{}l@{}}Recommendation of artifacts that \\ should be edited according to the \\ project history captured.\end{tabular} &  \\ \hline
\textbf{SBA1} & \begin{tabular}[c]{@{}l@{}}Executing IDE commands and verification if \\ contextual factors of the proposed model \\ correlate with commands in the IDE.\end{tabular} & \begin{tabular}[c]{@{}l@{}}Concerned with the \\ privacy of developers, \\ the work is limited to \\ IDE commands.\end{tabular} & \begin{tabular}[c]{@{}l@{}}Advantages: The context modeled provide \\ meaningful information regarding the \\ interactions with the IDE while developing. \\ Disad: there are no evaluations considering \\ if the performance of the developers \\ improved, for example.\end{tabular} & \begin{tabular}[c]{@{}l@{}}During evaluation, the model was \\ populated with information, so \\ statistical analysis was possible.\end{tabular} & \begin{tabular}[c]{@{}l@{}}There are contextual\\ factors abstracted to \\ represent an artifact.\end{tabular} \\ \hline
\textbf{SBA2} & Inteview and experiment mixed. &  & \begin{tabular}[c]{@{}l@{}}Adv: Allows reduced context switches \\ through natural language.\end{tabular} &  & \begin{tabular}[c]{@{}l@{}}The speech is abstracted \\ into the set of contexts \\ expected by the tool.\end{tabular} \\ \hline
\textbf{SBA3} &  &  &  &  &  \\ \hline
\textbf{SBA4} & \begin{tabular}[c]{@{}l@{}}Evaluated ranking of Stack \\ Overflow posts and the usefulness \\ of the tool proposed.\end{tabular} &  & \begin{tabular}[c]{@{}l@{}}Adv.: Supports provision of flow to  \\ software development as developers \\ do not need to leave the IDE to \\ search for support when coding.\end{tabular} &  &  \\ \hline
\textbf{SBA5} &  & \begin{tabular}[c]{@{}l@{}}Suggestions can take \\ long in case of \\ modified code\end{tabular} &  &  & \begin{tabular}[c]{@{}l@{}}Users are able to set tags \\ (concerns) which the \\ proposed algorithm \\ also uses when trying to \\ find the related contexts\end{tabular} \\ \hline
\textbf{SBA6} & \begin{tabular}[c]{@{}l@{}}Qualitative evaluation on the usefulness of the \\ recommendations. Results show \\ recommendations were helpful.\end{tabular} &  &  &  &  \\ \hline
\textbf{SBA7} & \begin{tabular}[c]{@{}l@{}}Quantitative and qualitative field study \\ with participants. Both produced\\ evidence that the use of task context can\\  make programmers more productive.\end{tabular} &  &  &  & \begin{tabular}[c]{@{}l@{}}Programmers tasks \\ are abstracted into \\ high-level tasks.\end{tabular} \\ \hline
\end{tabular}
\end{table}
\end{landscape}

\end{document}